\documentstyle[preprint,aps,eqsecnum]{revtex}
\begin {document}
\preprint{OITS-578}

\draft

\title{MULTIPLICITY DISTRIBUTIONS OF \\ SQUEEZED ISOSPIN STATES }

\author{I.\ M.\ Dremin\footnote{Permanent address:  P.\ N.\ Lebedev
Physical Institute, Leninsky prospect, 53, Moscow 117924 Russia.  Electronic
address:  dremin@lpi.ac.ru } and Rudolph C.\ Hwa\footnote{Electronic
address:  hwa@oregon.uoregon.edu}}
\address{ Institute of Theoretical Science and Department of  Physics,\\
University of Oregon, Eugene, Oregon 97403-5203}
\maketitle
\begin{abstract}
 Multiplicity distributions of neutral and charged particles arising from
squeezed coherent states are investigated.  Projections onto global isospin
states are considered.  We show how a small amount of squeezing can
significantly change the multiplicity distributions.  The formalism is
proposed to describe the phenomenological properties of neutral and
charged particles anomalously produced in hadronic and nuclear collisions
at very high energies.
\end{abstract}
\newpage
\section{Introduction}

There has been a long-standing interest in the study of the multiplicity
distributions of coherent states modified by various dynamical constraints
\cite{1,2,3,4,4.5,5}.  The recent interest in the disoriented chiral condensate
(DCC)
\cite{6} as a possible explanation of the Centauro-type events \cite{7} has
further extended the use of coherent states in the study of unusual charge
distributions \cite{8}.  It has also been suggested \cite{9,10} that a natural
quantum description of the DCC can be given in terms of the squeezed  states
\cite{11,12,13,14}.  In this paper we investigate the neutral and charge
distributions of squeezed  coherent states constrained to form global isospin
scalar and vector states that can possibly arise in nuclear collisions at very
high energies.  Our emphasis will be on the multiplicity distributions of the
charged and neutral pions.  It is therefore a significantly more detailed
investigation than would be required, if the goal is only to determine the
average fraction of neutral particles produced.

If charged pions are always produced in pairs to maintain local charge
neutrality of the source medium, it is very natural to consider squeezed
states, which are created from the vacuum by operators that are
exponentiations of the quadratic forms of the creation and annihilation
operators.  Although the squeezed  states are constructed in quantum optics
to reduce noise in certain applications
\cite{11,12,13,14}, their mathematical structure has features that are useful
in quantizing classical fields.  Their relevance to hadron production in
high-energy collisions was recognized long ago
\cite{4}.  Recent interest in the subject has been stimulated by the study of
chiral phase transition, where the long wavelength pion mode is found in the
linear
$\sigma$ model to grow rapidly after quenching \cite{15,16}.  It need not be
a DCC, but the interest in developing a quantum description of DCC has led to
the realization of the relevance of the squeezed  states \cite{10}.

In our view that is rooted in the phenomenology of hadronic and nuclear
collisions, the likelihood of the existence of a long wavelength pion mode or
the creation of a DCC shielded from the normal vacuum by a hot shell in a
heavy-ion collision is very remote, no matter how ideal or contrived the
collision process may be.  While we can think of various arguments against
the conditions favorable for their occurrence, there has not been any serious
effort made to present a phenomenologically reasonable scenario in the form
of realistic modeling that can demonstrate the feasibility of their appearance
in actual collisions.  Thus the subject is driven so far mainly by the
attractiveness of the theoretical idea and the alluring possibility of solving
concrete classical equations.  Until such time as when experiments suggest
the relevance of disoriented chiral condensates to the anomalous production
of charge and neutral particles, we shall regard the subject as an interesting
speculation that does not preempt other possible approaches.

We begin with the observation of two facts.  One is the empirical fact that
anomalous events like the Centauro have been recorded in cosmic-ray
experiments\cite{7}, though not in the laboratory.  The other is the lack of a
theoretical framework to describe the unusual charge distributions
independent of the chiral dynamics.  It seems to us useful to have a language
to convey quantitatively the characteristics of the distributions without a
rigid dynamical basis so that one may have what amounts to a representation
in much the same way that the negative binomial distributions have been
used to describe multiplicity distributions in general \cite{4,17,18}.  While
there exist earlier attempts to describe exotic neutral and charge
distributions in terms of the coherent states \cite{2,3}, it is better for our
purpose to introduce some parameters that can be adjusted to change the
neutral-to-charge ratio.  To that end we find that the squeezed  coherent
states are well endowed with features suitable for our needs.  Moreover, the
consideration of global isotopic spin turns out to be crucial, as in \cite{1},
since unlike quantum optics the squeezed  coherent states produced in
hadronic and nuclear collisions must have a limited range of possible initial
isotopic spin states.  We should emphasize, however, that the dynamical
origin of the squeezed  coherent states in such collisions is not an issue that
must be addressed first; we use the squeezed coherent state here only as a
mathematical framework to discuss the production of neutral and charge
quanta that are locally neutral in a certain limit (squeezed vacuum).

We give in Sec. II a brief review of the squeezed  coherent state.  Application
to particle production is considered in Sec. III, and the results on
multiplicity
distributions are presented in Sec. IV.
\section{Squeezed  States} In this section we summarize the basic properties
of squeezed states
\cite{11,12,13,14,19}  that are to be used in the following section.

We start with the coherent state $| \alpha \rangle$, created from the
vacuum by the displacement operator $D(\alpha)$:
\begin{eqnarray}
 |\alpha \rangle = D(\alpha) | 0 \rangle
\label{2.1}
\end{eqnarray}
where
\begin{eqnarray}
 D(\alpha) = e ^{\alpha a^{\dagger} - \alpha^{\ast}a} \quad .
\label{2.2}
\end{eqnarray}
$a^{\dagger}$ and $a$ are creation and annihilation operators, and $\alpha$
is a complex number.  From (\ref{2.2}) we have
\begin{eqnarray}
D^{\dagger}(\alpha)\, a\, D(\alpha) = a + \alpha \quad ,
\label{2.3}
\end{eqnarray} whereupon we obtain the fundamental property that
$|\alpha \rangle$, defined in (\ref{2.1}), is an eigenstate of $a$
\begin{eqnarray}
 a|\alpha\rangle = \alpha|\alpha\rangle
\label{2.4}
\end{eqnarray} with eigenvalue $\alpha$.

The squeeze operator $S(\eta)$ is defined by
\begin{eqnarray}
S(\eta) = \mbox{exp} \left[ (\eta ^{\ast}a^2 - \eta a
^{\dagger 2})/2\right]
\label{2.5}
\end{eqnarray}
where $\eta = re^{i2\psi}$.  A unitary transformation on $a$ by
$S(\eta)$ yields
\begin{eqnarray}
 S^{\dagger}(\eta)\,a\, S(\eta) = \mu a - \nu a ^{\dagger}
\quad ,
\label{2.6}
\end{eqnarray} where
\begin{eqnarray}
\mu = \mbox {cosh}\, r, \qquad \nu = e^{i2\psi} \mbox {sinh}\, r \quad .
\label{2.7}
\end{eqnarray}
Inverting (\ref{2.6}) and a similar one on  $a^{\dagger}$ gives
\begin{eqnarray}
 a(\eta) \equiv S(\eta)\,a \,S^{\dagger}(\eta) = \mu \, a +
\nu \, a^{\dagger} \quad .
\label{2.8}
\end{eqnarray}
The unitary transformation of $S(\eta)$ on the displacement
operator yields
\begin{eqnarray}
 S^{\dagger}(\eta)D(\alpha)S(\eta) = D(\alpha(\eta))\quad ,
\label{2.9}
\end{eqnarray}
where
\begin{eqnarray}
\alpha(\eta) = \mu \alpha + \nu \alpha ^{\ast} \quad .
\label{2.10}
\end{eqnarray}

The squeeze state $|\alpha, \eta \rangle$ is defined by
\begin{eqnarray}
|\alpha, \eta \rangle = D(\alpha)\, S(\eta) | 0 \rangle \quad.
\label{2.11}
\end{eqnarray}
Evidently, when $\eta = 0$, i.e. no squeezing, $|\alpha,
0\rangle$ is identical to the coherent state $|\alpha \rangle$.  Even for $\eta
\neq 0$, the expectation value of $a$ is still $\alpha$, since we have, using
(\ref{2.3}) and (\ref{2.6})
\begin{eqnarray}
\left<\alpha, \eta|\,a\, |\alpha, \eta\right\rangle = \left<0|
S^{\dagger}(\eta)
(a +
\alpha)S(\eta)|0\right\rangle = \alpha \quad .
\label{2.12}
\end{eqnarray}
However, $|\alpha, \eta\rangle$ is not an eigenstate of $a$
for
$\eta \neq 0$; instead, it is  an eigenstate of $a(\eta)$, defined in
(\ref{2.8}),
i.e.,
\begin{eqnarray}
a(\eta)|\alpha, \eta\rangle &=& S(\eta) a
S^{\dagger}(\eta)\,D(\alpha)S(\eta) |0 \rangle\nonumber \\
 &=& S(\eta) a |\alpha(\eta)\,\rangle = \alpha(\eta)|\alpha,\eta\,\rangle
\quad .
\label{2.13}
\end{eqnarray}
 Inverting (\ref{2.8}) we have
\begin{eqnarray} a = \mu a (\eta) - \nu a^{\dagger} (\eta)
\label{2.14}
\end{eqnarray}
so that
\begin{eqnarray}
a^{\dagger}a = \left(\mu ^2 + |\nu|^2 \right) a^{\dagger}
(\eta) a(\eta)\, -\, \mu
\nu a^{\dagger}(\eta)^2 \, -\, \mu
\nu ^{\ast} a(\eta)^2 + |\nu|^2 \quad ,
\label{2.15}
\end{eqnarray}
where $\left[ a(\eta), a^{\dagger}(\eta)\right] = 1$ has been
used.  It then follows from (\ref{2.13}) and (\ref{2.10}) that
\begin{eqnarray}
\left< \alpha, \eta | a^{\dagger} a |\alpha, \eta\right> = |\alpha|^2 + |\nu|^2
\quad.
\label{2.16}
\end{eqnarray}
Thus the expectation value of the number operator for the
squeezed state differs from that for the pure coherent state by the additive
term $|\nu|^2$.

\section{Multiplicity Distributions for Squeezed Isospin States}

We now generalize the squeezed states described in the previous section to
the isospin space in a manner similar to how the coherent states are
generalized in Ref. 1.  The annihilation operator $\vec{a}$ is now a vector in
the isospin space, and the corresponding eigenstate $|\vec{\alpha}\rangle$ is
created from the vacuum by the operator  $D (\vec{\alpha})$.  The squeezed
state is then
\begin{eqnarray}
 | \vec{\alpha}, \eta \rangle\,  = D (\vec{\alpha}) \, S (\eta)
| 0 \rangle
\quad.
\label{3.1}
\end{eqnarray}
The complex vector $\vec{\alpha} = \alpha \hat{n}$ has a
direction in the isospin space, specified by the polar angles $\theta$ and
$\phi$.  A squeezed state with a specific total isospin $I$ and $z$-component
$I_z$ can then be expressed in terms of $| \vec{\alpha}, \eta \rangle$ in the
usual way
\begin{eqnarray}
|\alpha, \eta \rangle _{I, I_z}\, = \int d \Omega_{\hat{n}}\,
Y^{\ast} _{I, I_z} \left(
\theta,
\phi \right) | \vec{\alpha}\left( \theta,
\phi \right), \eta \rangle \quad .
\label{3.2}
\end{eqnarray} What we need to do is to determine the projection of this
state to the eigenstates of the number operators specifying the neutral
$(n_0)$ and charged $(n_c)$ particle multiplicities.

Then the multiplicity distribution of a squeezed isospin state is
\begin{eqnarray}
 P^{I, I_z}_{n_0, n_c} \left( \alpha, \eta \right) \, = \left|
\left< n_0, n_c|\alpha, \eta\right>_{I, I_z}\right|^2 \quad .
\label{3.3}
\end{eqnarray}
It will be parameterized by the variables $\alpha$ and
$\eta$.  More precisely, the results will depend only on the magnitudes
$|\alpha|$ and $r$, as one might surmise from (\ref{2.16}).

To proceed, it is necessary to determine first $\left< n_0, n_c|\vec{\alpha},
\eta\right>$, which in turn can follow from a knowledge of the scalar product
between a squeezed state and a coherent state.  Following Yuen
\cite{11}, we first note that the differential operator representation of any
polynomial $M\left( \vec{a},\vec{a}^{\dagger}\right)$ of $\vec{a}$ and
$\vec{a}^{\dagger}$ in the coherent-state representation is
\begin{eqnarray}
\left<\vec{\beta}\left| M\left( \vec{a},\vec{a}^{\dagger}\right)\right| \psi
\right> = M \left( {\vec{\beta} \over 2} + {\partial \over \partial
\vec{\beta}^{\ast}}\, ,
\vec{\beta}^{\ast}\right) \left< \vec{\beta}|\psi \right> \quad .
\label{3.4}
\end{eqnarray} Applying this to $\vec{a}(\eta)$ for $M\left(
\vec{a},\vec{a}^{\dagger}\right)$, we have with the help of (\ref{2.13})
\begin{eqnarray}
\left<\vec{\beta} |\vec{a}(\eta)|\vec{\alpha}, \eta\right> = \vec{\alpha}(\eta)
\left< \vec{\beta} | \vec{\alpha}, \eta\right> = \left[ \mu  \left(
{\vec{\beta}
\over 2} + {\partial \over \partial
\vec{\beta}^{\ast}}\right)+\nu\vec{\beta}^{\ast}\right]\left<\vec{\beta}
|\vec{\alpha}, \eta\right>\quad.
\label{3.5}
\end{eqnarray} This is a differential equation on $\left<
\vec{\beta}|\vec{\alpha},\eta
\right>$ that has the solution
\begin{eqnarray}
\left<\vec{\beta} |\vec{\alpha}, \eta\right> = \mbox{exp} \left[ - {1 \over
2}\left|\vec{\beta}\right|^2 - {\nu \over 2 \mu} \vec{\beta}^{\ast} \cdot
\vec{\beta}^{\ast} + {1 \over
\mu} \vec{\beta}^{\ast} \cdot \vec{\alpha} (\eta) + f \left(\vec{\alpha}
(\eta),
\vec{\alpha}^{\ast} (\eta)
\right)\right]\quad.
\label{3.6}
\end{eqnarray} The last term that depends only on $\vec{\alpha} (\eta)$ and
$\vec{\alpha}^{\ast} (\eta)$ can be determined by working with
\begin{eqnarray}
\left< \vec{\alpha}, \eta|\vec{a}(\eta)|\vec{\beta}\right> &=& \left[
{\vec{\alpha}(\eta)\over 2} {\partial \over \partial \vec{\alpha}^{\ast}
(\eta)}\right] \left< \vec{\alpha}, \eta|\vec{\beta}\right> \nonumber\\
 &=& \left( \mu \vec{\beta}  + {\nu \over 2}\vec{\beta}^{\ast} + \nu {\partial
\over \partial \vec{\beta}} \right)
\left<
\vec{\alpha},
\eta|\vec{\beta}\right> \quad .
\label{3.7}
\end{eqnarray}
The required form for $f\left(\vec{\alpha} (\eta),
\vec{\alpha}^{\ast} (\eta)\right)$ is therefore
\begin{eqnarray}
f\left( \vec{\alpha}(\eta), \vec{\alpha}^{\ast}(\eta)\right) =
- { 1\over 2}
\left|\vec{\alpha}(\eta)\right|^2 + {\nu ^{\ast} \over 2 \mu} \vec{\alpha}^2
(\eta)
\quad .
\label{3.8}
\end{eqnarray}

Since $\vec{\beta}$ represents both the neutral and charged
sectors, $\beta \,
\mbox{cos} \theta$ and $\beta \, \mbox{sin} \theta$, in the isospin space,
respectively, we have the usual representation of a coherent state
$|\vec{\beta}\rangle$ in the $| n_0, n_c \rangle$ basis, i.e.,
\begin{eqnarray}
\left< n_0, n_c | \vec{\beta} \right> = {\beta^{n_0 +  n_c} \over \sqrt{n_0!
n_c!}}
\left(\mbox{cos}\theta\right)^{n_0}  \left(\mbox{sin}\theta\right)^{n_c}
e^{-\left|\vec{\beta}\right|^2/2}\Phi(\phi)\quad,
\label{3.9}
\end{eqnarray}
where $\Phi (\phi) = \mbox{exp}\left[
i(n_+-n_-)\phi\right]$ is a phase factor that depends on the difference
between $+$ and $-$ charged-particle multiplicities.

Using (\ref{3.6}), (\ref{3.8}) and (\ref{3.9}) on
\begin{eqnarray}
\left<\vec{\beta} |\vec{\alpha}, \eta\right> = \sum_{n_0, n_c}
\left<\vec{\beta} | n_0, n_c
\right>\left< n_0, n_c | \vec{\alpha},\eta \right>
\label{3.10}
\end{eqnarray}
and the expansion
\begin{eqnarray}
e^{2\vec{z}\cdot\vec{t}-\vec{t}^2} = \sum_{n_0, n_c}
{t_0^{n_0}t_c^{n_c} \over n_0! n_c!} H_{n_0} (z_0) H_{n_c} (z_c)\quad ,
\label{3.11}
\end{eqnarray}
where $t_0 = t \, \mbox{cos}\theta, t_c = t \, \mbox{sin}\theta$
and similarly for
$z_0$ and $z_c$, $H_n (z)$ being the Hermite polynomial, we have
\begin{eqnarray}
\left< n_0, n_c | \vec{\alpha},\eta \right> = A_{n_0, n_c} \left[\alpha (\eta)
\right] H_{n_0}\left[ {\alpha_0(\eta) \over \sqrt{2 \mu \nu}}\right] \
H_{n_c}\left[
{\alpha_c(\eta)
\over \sqrt{2 \mu \nu}}\right] \Phi(\phi) \quad ,
\label{3.12}
\end{eqnarray}
where
\begin{eqnarray} A_{n_0, n_c} \left[\alpha (\eta) \right] = (n_0! n_c!)^{-{1
\over 2}} \left( {\nu
\over 2 \mu}\right)^{(n_0 + n_c)/2} \mbox{exp} \left[ - {1 \over 2}
\left|\vec{\alpha}(\eta)\right|^2 +  {\nu^{\ast}
\over 2 \mu} \vec{\alpha}^2(\eta)\right] \quad .
\label{3.13}
\end{eqnarray}
Note that $A_{n_0, n_c} \left[\alpha (\eta) \right]$ is
independent of the isospin direction, whereas the arguments of $H_n$ are,
since $\alpha _0(\eta) = \alpha(\eta) \mbox{cos}\theta$ and $\alpha
_c(\eta) = \alpha(\eta) \mbox{sin}\theta$.  For the isospin
projected state (\ref{3.2}), we obtain correspondingly
\begin{eqnarray}
\left< n_0, n_c | \alpha,\eta \right>_{I,I_z} = A_{n_0, n_c} \left(\alpha, \mu,
\nu
\right) B^{I,I_z}_{n_0, n_c} (b)\quad ,
\label{3.14}
\end{eqnarray}
where
\begin{eqnarray} B^{I,I_z}_{n_0, n_c} (b) = \int^1_{-1} d \mbox{cos}\theta
\int^{2\pi}_0 d \phi Y^{\ast}_{I,I_z}(\theta,\phi) H_{n_0} (b \,
\mbox{cos}\theta) H_{n_c} (b \,
\mbox{sin}\theta) \Phi(\phi) \quad ,
\label{3.15}
\end{eqnarray}

\begin{eqnarray}
b = \alpha (\eta)/ \sqrt{2\mu \nu} \quad .
\label{3.16}
\end{eqnarray}
When used in (\ref{3.3}), this gives us the multiplicity
distributions that we seek.

We have not given any special attention to the normalization of
$\left< \vec{\beta}|\vec{\alpha}(\eta)\right>$ as a solution to the
differential
equation in (\ref{3.5}).  That freedom will be used in normalizing
$P^{I,I_z}_{n_0, n_c} (\alpha, \eta)$ at the end.

There is more freedom in the squeezed states than is needed in our problem.
{}From (\ref{2.7}) and (\ref{2.10}) we recall that
\begin{eqnarray}
\vec{\alpha}(\eta) = \left( \alpha \,\mbox{cosh}\, r\,+ \alpha ^{\ast}
\mbox{sinh}\, r\, e^{i2\psi}\right)\hat{n}\quad.
\label{3.17}
\end{eqnarray}
Let us set $\alpha, \mu,$ and $\nu$ all real so that
\begin{eqnarray}
\vec{\alpha}(\eta) = \alpha (\mu + \nu) \hat{n} \quad ,
\label{3.18}
\end{eqnarray}
and
\begin{eqnarray}
b = {\alpha (\mu + \nu)  \over  \sqrt{2 \mu \nu}} \quad ,
\quad \mu = \sqrt{1 +
\nu ^2}  \quad .
\label{3.19}
\end{eqnarray}
{}From (\ref{3.3}) and (\ref{3.14}) we have
\begin{eqnarray}
P^{I,I_z}_{n_0, n_c} (\alpha, \nu) = {N \over n_0! n_c! }
\left({\nu \over 2\sqrt{1 +
\nu ^2}} \right)^{n_0 + n_c} \left|  \, B^{I,I_z}_{n_0, n_c} (b)\,\right|^2
\label{3.20}\quad ,
\end{eqnarray}
where $N$ is a normalization factor.

The average multiplicity $\langle n\rangle$ is
\begin{eqnarray}
\left< n\right> = \left< n_0\right> + \left<n_c\right>  =
\alpha ^2 +
\nu ^2
\quad ,
\label{3.21}
\end{eqnarray}
 as required by (\ref{2.16}), which is invariant under rotation
in the isospin space.  We shall find it more convenient in the following to use
the squeeze parameter $s$, defined by
\begin{eqnarray}
s = (\nu / \alpha) ^2
\label{3.22}
\end{eqnarray}
so that
\begin{eqnarray}
\left< n\right> = \alpha ^2 (1 + s) \quad .
\label{3.23}
\end{eqnarray}
In all instances that we shall consider below, the value of $s$
will be small, so $\langle n\rangle$ will not be effected very much by
squeezing.
Nevertheless, we shall see that for certain isospin states the $n_0$ and $n_c$
distributions can be significantly influenced by the small amount of
squeezing.

In the next section we shall calculate $P^{I,I_z}_{n_0, n_c}$ only for $I = 0$
and $1$, although there exist no obstacles in carrying out the computation at
higher values of $I$.  For $p\bar p$ annihilation $I \leq 1$ is all that is
necessary.  For $pp$ inelastic collisions the isospin of the pion state can in
principle be as large as $I=2$, but we shall restrict our consideration here to
only $I \leq 1$.  It is tacitly assumed that the analysis considered here is
only appropriate for the particles produced in the central region.  Even for
nuclear collisions we hope that $I \leq 1$ is sufficient.

For $I = 0, 1$, we can write
\begin{eqnarray}
Y_{I,I_z}(\theta, \phi) = c_{I,I_z}
\left(\mbox{cos}\theta\right)^{I-|I_z|} \left(\mbox{sin}\theta\right)^{|I_z|}
e^{iI_z\phi}
\quad .
\label{3.24}
\end{eqnarray}
Thus the $\phi$ integration in (\ref{3.15}) requires $n_+-n_-
= I_z$.  For the
$\theta$ integration let us define
\begin{eqnarray}
 J^{k, l}_{n_0, n_c} (b) = \int^1_{-1} dx\, x^k (1 - x^2)^{l/2}
H_{n_0}(bx) H_{n_c} (b\sqrt{1-x^2})
\label{3.25}
\end{eqnarray}
so that $B^{I,I_z}_{n_0, n_c} (b)$ is proportional to it.  Let $G
(t_1, t_2)$ be the generating function, from which $J^{k, l}_{n_0, n_c}$ can be
obtained:
\begin{eqnarray}
 J^{k, l}_{n_0, n_c} (b) = \left.{{\partial^{n_0} \over \partial
t^{n_0}_1} {\partial^{n_c} \over
\partial t^{n_c}_2} G (t_1, t_2)}\right| _{t_1= t_2 =0}
\quad .
\label{3.26}
\end{eqnarray}
Eq. (\ref{3.25}) implies
\begin{eqnarray}
G (t_1, t_2) &=& \sum _{n_0, n_c} \int^1_{-1} dx\, x^k (1 -
x^2)^{l/2} H_{n_0} (bx) H_{n_c} \left(b\sqrt{1-x^2}\right) {t^{n_0}_1 t^{n_c}_2
\over  n_0! n_c!}\nonumber\\
&=& e^{-t^2_1 -t^2_2 } \int^1_{-1} dx\, x^k (1
- x^2)^{l/2} e^{2b \left(xt_1 +
\sqrt{1 - x^2} t_2\right)} \nonumber\\
 &=& e^{-t^2_1 -t^2_2 }\sum ^{\infty}_{n,m=0} (2b) ^{n+m}\, {t^n_1 t^m_2
\over  n! m!}\, B \left( {n + k + 1 \over 2}, {m + l + 2 \over 2}\right)
\epsilon_{n + k} \quad ,
\label{3.27}
\end{eqnarray}
where $B(u, v)$ is the Euler beta function.  The factor
$\epsilon_{n+k}$ is $1$ if
$n+k$ = even, $0$ if $n+k$ = odd; it is due to the symmetry of the integrand
under $x \leftrightarrow -x$.  It can be shown after some algebraic
manipulations that
\begin{eqnarray}
\left.{{d^q \over dt^q} \left(e^{-t^2} t^n \right)}\right|_{t = 0} =
(-1)^{q-n} { q! \over (q-n)!} H_{q-n}(0) \quad .
\label{3.28}
\end{eqnarray}
Since the order of $p$ of $H_p(z)$ must be even when $z =
0$, $q - n$ must be even, so $(-1)^{q-n} = 1$.  Using this in (\ref{3.26}) and
(\ref{3.27}) yields
\begin{eqnarray}
J^{k, l}_{n_0, n_c} (b) &=& \sum ^{n_0}_{n=0}\sum
^{n_c}_{m=0} (2b) ^{n+m}  H_{n_0-n}(0) H_{n_c-m}(0)
{n_0 \choose n}
{n_c\choose m}\\ \nonumber\
&&B
\left( {n + k + 1
\over 2}, {m + l + 2
\over 2}\right) \epsilon_{n + k} \quad.
\label{3.29}
\end{eqnarray}
We finally have from (\ref{3.20})
\begin{eqnarray}
 P^{I,I_z}_{n_0, n_c} (\alpha, s) = {N^{\prime} \over n_0! n_c!
} \left[4 \left(1 + {1
\over s \alpha ^2}\right) \right] ^{-\left(n_0+ n_c\right)/2}\left|
J^{k,l}_{n_0,
n_c}(b)\right|^2
\label{3.30}\quad ,
\end{eqnarray}
where $N^{\prime}$ is another normalization constant.  From
the exponents in (\ref{3.24}) and (\ref{3.25}), we have

\begin{eqnarray}
\begin{array}{lll}
 I = 0:& k = 0,\, l = 0, &n_+ = n_- \quad.\\
\label{3.31a} I = 1, I_z = 0:\hspace{.5cm}& k = 1, \,l = 0, \hspace{.5cm}&n_+ =
n_- \quad.\\
\label{3.31b}
 I = 1, I_z = 1:&k = 0,\, l = 1, &n_+ = n_-+1 \quad.
\label{3.31c}
\end{array}
\end{eqnarray}
Clearly, $n_c = n_++n_-$ is even (odd), when $n_+ = n_- \ \,(n_-+1)$.
When
$k = 0$, $n$ must be even, so $n_0$ must also be even; for $k = 1$, $n_0$ is
odd.  Thus the evenness or oddness of $n_0$ and $n_c$ can be summarized
as follows:
\begin{eqnarray}
\begin{array}{llll}
 \mbox{(a)}\qquad \qquad & I = 0: &n_0 = \mbox{even}, &n_c =
\mbox{even}
\quad.
\\
\label{3.32a}
 \mbox{(b)}\hspace{.25cm}& I = 1, I_z = 0: \hspace{.5cm}&n_0 =
\mbox{odd},
\hspace{.5cm}&n_c =
\mbox{even}
\quad.\\
\label{3.32b}
 \mbox{(c)}& I = 1, I_z = 1:& n_0 = \mbox{even}, &n_c =
\mbox{odd}
\quad.
\label{3.32c}
\end{array}
\end{eqnarray}

\section{Results on Multiplicity Distributions}

In this section we present the results of our calculation of $P^{I,I_z}_{n_0,
n_c}
(\alpha, s)$.  Our objective is to provide some insight into the dependences
on squeezing and on the isospin states.  To that end we shall fix the average
multiplicity $\left<n\right>$ at some reasonable value, which we take to be
25.  Thus
$\alpha$ and $s$ are constrained through (\ref{3.23}).  To demonstrate the
sensitivity on squeezing for certain isospin states, we shall consider only two
small values of $s$, {\it viz.}, $s = 0.02$ and $0.04$, which is sufficient to
reveal the qualitative features of the $s$ dependence.  Other higher values of
$s$ can readily be considered, but are not examined here to avoid
visual confusion in the results to be presented.

In calculating $P^{I,I_z}_{n_0, n_c} (\alpha, s)$ we note that $s$, as defined
in
(\ref{3.22}), is independent of the sign of $\nu$.  In (\ref{3.19}) we have
taken $\mu$  and $\nu$ to be real, but it does not preclude $\nu$ to be
negative, corresponding to $\psi = \pi /2$ in (\ref{2.7}).  Since the result
will
depend on the sign of $\nu$, we shall use the notation $s_{\pm}$, to stand
for sgn$(\nu) = \pm$, while specifying the value of $s$.  In the following the
distributions of the neutral and charged sectors are given separately, where
for fixed $\left<n\right>$
\begin{eqnarray} P^{I,I_z}_{n_0} (s_{\pm}) = \sum_{n_c} P^{I,I_z}_{n_0, n_c}
\left(\alpha, s_{\pm}\right)
\quad ,
\label{4.1}
\end{eqnarray}
\begin{eqnarray} P^{I,I_z}_{ n_c}(s_{\pm}) = \sum_{n_0} P^{I,I_z}_{n_0, n_c}
\left(\alpha, s_{\pm}\right)
\quad.
\label{4.2}
\end{eqnarray}

For $I = 0$, the $n_0 $ distribution
$P^{0}_{n_0}\left(s_{\pm}\right)$ is insensitive to $s_{\pm}$, but the
$P^{0}_{n_c}\left(s_{\pm}\right)$ is highly sensitive to $s_{\pm}$.  This is
shown in Figs. 1 and 2 for the cases $\nu > 0$ and  $\nu < 0$, respectively.
Clearly,  $P^{0}_{n_0}$ is peaked at low
$n_0$, while the peak of $P^{0}_{n_c}$ shifts to low $n_c$ with increasing
$s_-$, although stationary in  $s_+$.  The behavior of $P_{n_0}^0$ for
squeezed states is very similar to the coherent case considered in \cite{3}.
There, the asymptotic behavior is $f^{-1/2}$, where $f=n_0/(n_0+n_c)$,
(obtained by using the Stirling formula in analytical estimates).  Thus that
law is not greatly influenced by slight squeezing.

For $I = 1$, $I_z = 0$,  $\nu > 0$, $P^{1,0}_{n_0}\left(s_{+}\right)$ is very
nearly the same as $P^{1,0}_{n_c}\left(s_{+}\right)$, even though the former
is only for odd $n_0$ and the latter for even $n_c$, as shown in Fig. 3. Their
shapes depend sensitively on $s_+$.  For $I = 1$, $I_z = 0$,  $\nu < 0$, the
distributions are very different; $P^{1,0}_{n_0}\left(s_{-}\right)$ is not
sensitive to $s_-$, but  $P^{1,0}_{n_c}\left(s_{-}\right)$ is.  That is
shown in Fig. 4.  Upon comparing these to  $P_{n_0,n_c}^0$,
it is evident that the distributions depend crucially on the isospin, even
though $I_z=0$ in both cases.

Finally, for $I = 1$, $I_z = 1$,  $P^{1,1}_{n_0}\left(s_{\pm}\right)$
behaves very similarly as $P^{0}_{n_0}\left(s_{\pm}\right)$, with peaking at
low $n_0$ and independence on $s_{\pm}$, as is clear in Figs. 5 and 6.
$P^{1,1}_{n_c}\left(s_{\pm}\right)$ is also similar to
$P^{0}_{n_c}\left(s_{\pm}\right)$, but the former is only for odd $n_c$, while
the latter is for even $n_c$.  Thus the $f^{-1/2}$ behavior is restored again.

It is relevant to ask whether the $n_t=n_0+n_c \to \infty$ limit should be
independent of squeezing.  There are two cases to consider:  $s \to 0$ and $s$
finite.   The $s$ dependence can be seen by examining (3.29).  For
infinitesimal  $s$,
$b$ is large, so the term with $n=n_0$ and $m=n_c$ is dominant.  Then the
large $n_t$ behavior is independent of the Hermite polynomials.  In the $n_t
\to \infty$ limit the $s$ dependence of (3.29) cancels that in (3.20),
resulting in no net dependence on $s$.  But if $s$ is set at a finite value
initially,  all terms in the sum in (3.29) can be important, since
$H_p(0)$ diverges with increasing $p$.  The large $n_t$ limit of
$J_{n_0,n_c}^{k,l}$ can then be very different and $P^{I,I_z}_{n_0, n_c}$ can
depend in $s$, even if $s$ is not large.  In both cases, if $n_t$ is not
infinite,
there would in general be some dependence on $s$, the details of which can
only be seen from the numerical results.  It is for this reason that our
results
depend on squeezing.

To summarize, we observe that in all isospin states the $n_c$ distributions
are most sensitive to squeezing, not just in the value of $s$, but especially
in
the sign of $\nu$.
That is to be contrasted from the peaking of the neutral distribution
$P_{n_0}$ at $n_0=0$ for both the $I=0$ state and the $I=1, I_z=1$ state,
revealing the preference for small numbers of neutral pions for either sign
of $\nu$. On the other hand, for $I=1, I_z=0$ the neutral to charge ratio is
about 1 when $\nu>0$, but is greater than 1 when $\nu<0$. The latter
therefore leans toward the characteristics of the anti-Centauro events.
 By
varying $I$, $I_z$, and $s_{\pm}$, one can obtain a wide variety of possible
shapes of $P_{n_c}$ that could be used to fit the observed charged particle
distribution in high-energy collision for any given average total multiplicity.
In that sense its usefulness is rather similar to that of the negative binomial
distribution, which has a $k$ parameter that adjusts the width.  It should be
noted that neither the negative binomial nor the pure squeezed-state
distributions can reproduce properly the behavior of the cumulants
observed in the experiments \cite{20,21}, whereas the question about
squeezed isospin states is so far open. Unlike the negative binomial
distribution, we now have a great deal more freedom to accommodate very
anomalous production processes, where the neutral to charge ratio need not
be 1:2.  Indeed,
$P_{n_0}$ can be peaked at
$n_0 = 0$, which is a feature that may render this description relevant to the
Centauro events.  By varying the squeeze parameter, one can not only adjust
the neutral to charge ratio, but also change the detailed $n_0$ and $n_c$
dependences.

We have not made any proposal in this paper to suggest that the dynamical
origin for the anomalous production of particles is related to the squeezed
states.  We have only presented a formalism for the phenomenological
description of unusual $P_{n_0}$ and $P_{n_c}$.  If in the future such a
formalism is found empirically to be well suited for the data, then perhaps
one may be motivated to look deeper into the question of whether squeezed
states are produced in certain collisions at high energy.

\subsection*{Acknowledgment}

We are grateful to H. Carmichael and M. Raymer for very helpful discussions
on squeezed states.  This work was supported in part by a NATO grant under
reference CRG 930025, by the Russian Fund for Fundamental Research
under grant No. 93-02-3815, and by the U.S. Department of Energy under
grant No. DE-FG06-91ER40637.

\newpage
\subsection*{Figure Captions}
\begin{description}
\item[Fig. 1]\quad Multiplicity distributions of neutral and charged particles
for $I = 0$ and $\nu > 0$.  Open symbols are for neutral particles and full
symbols are for charged particles.  Circles (open and full) are for $s = 0.02$;
triangles (open and full) are for $s = 0.04$.

\item[Fig. 2] \quad Same as in Fig. 1 but for $I = 0$, $\nu < 0$.

\item[Fig. 3] \quad Same as in Fig. 1 but for $I = 1$, $I_z = 0$, $\nu > 0$.
(a)
and (b) are for both $n_0$ and $n_c$, but (c) is for $n_c$ only to show the
dependence on $s$.

\item[Fig. 4] \quad Same as in Fig. 3 but for  $I = 1$, $I_z = 0$, $\nu < 0$.

\item[Fig. 5] \quad Same as in Fig. 1 but for  $I = 1$, $I_z = 1$, $\nu > 0$.

\item[Fig. 6] \quad Same as in Fig. 1 but for  $I = 1$, $I_z = 1$, $\nu < 0$.

\end{description}

\end{document}